\documentclass[useAMS,usegraphicx]{mn2e}

\title[A z=3.045 Ly$\alpha$ emitting halo]{A z=3.045 Ly$\alpha$ emitting halo hosting a QSO and a possible candidate for AGN-triggered star-formation
 \thanks{This paper is based on data gathered with the 6.5 meter Magellan Telescopes located at Las Campanas Observatory, Chile.}}
\author[Michael Rauch et al.]{Michael Rauch$^{1}$, George D. Becker$^{2}$, Martin G. Haehnelt,$^{2}$
\newauthor Robert F. Carswell$^{2}$, Jean-Rene Gauthier$^{3}$ \\
$^{1}$Carnegie Observatories, 813 Santa Barbara Street, Pasadena, CA 91101, USA\\
$^{2}$Institute of Astronomy and Kavli Institute for Cosmology, Cambridge University, Madingley Road,  Cambridge CB30HA, UK\\
$^{3}$California Institute of Technology, Pasadena, CA 91125, USA}

\begin{document}


\pagerange{\pageref{firstpage}--\pageref{lastpage}} \pubyear{2011}

\maketitle


\label{firstpage}

\begin{abstract}  In this third paper in a series on the nature of extended, asymmetric Ly$\alpha$ emitters at $z\sim 3$ we report the discovery, in an ultra-deep, blind, spectroscopic long-slit survey,  of a Ly$\alpha$ emitting halo around a QSO at redshift 3.045.  The QSO is a previously known, obscured
AGN.  The Ly$\alpha$ emitting halo appears extended  along the direction of the slit and exhibits two faint patches separated
by 17 proper kpc in projection from the QSO.
Comparison of the 2-dimensional
spectrum with archival HST ACS images shows that these patches coincide spatially with emission from a peculiar, dumbbell-shaped, faint galaxy. 
The assumptions that the Ly$\alpha$ emission patches are originating in the galaxy and that the galaxy is physically related to the QSO
are at variance with photometric estimates of the galaxy redshift.
We show, however, that a population of very young stars at the redshift of the QSO may fit the existing
rest frame broad band UV photometry of the galaxy. If this scenario is correct then the symmetry of the galaxy in continuum and Ly$\alpha$ emission, the extension of the QSO's Ly$\alpha$ emission in its direction, and the likely presence of a young stellar population in close proximity to a (short-lived) AGN suggest that this may be an example of AGN feedback triggering external star formation in high redshift galaxies.

 \end{abstract}


\section[]{Introduction}

\begin{figure*}
\includegraphics[scale=.65,angle=0,keepaspectratio = true]{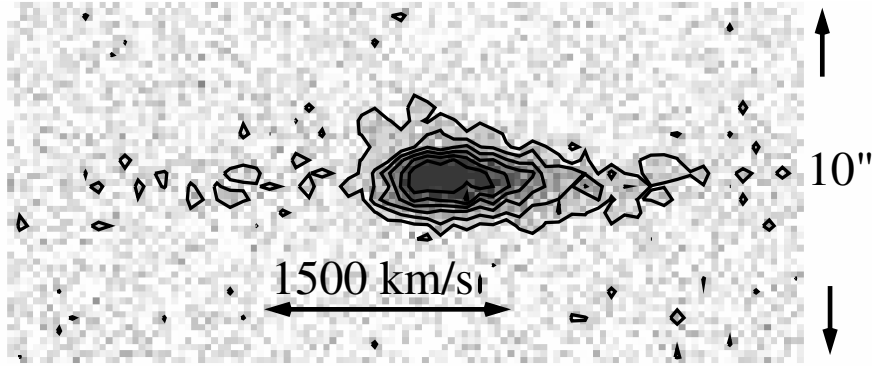}
\caption{LDSS3 spectrum of the QSO's Ly$\alpha$ emission line region. Flux density contours are plotted on top of the flux density spectrum, to give an idea of the noise and 
the dynamic range of the data. Wavelength (or redshift) increases from left to right. The direction along the slit is precisely N-S, with N on top. The outermost contiguous contour corresponds to a flux density of $7\times 10^{-20}$ erg cm$^{-2}$s$^{-1}$\AA$^{-1}$. The largest distance
of this contour from the (very faint) QSO continuum along the slit is about 2.2", or 17 proper kpc.\label{speccontour}}
\end{figure*}

Lyman $\alpha$ emitting halos around QSOs are a well documented and theoretically expected (e.g., Haiman \& Rees 2001) consequence of the enhanced ionization  of the surrounding gas by the central AGN. In contrast, the impact of AGN on star-formation in the QSO host galaxy or its satellite galaxies is more complex and subject to ongoing debate.
Most recently, the study of AGN feedback has  focused on the quenching of star formation (e.g., Silk \& Rees 1998; Fabian 1999; Schawinsky et al 2009;
Farrah et al 2012; Maiolino et a 2012). However,
considerable observational evidence (e.g., van Breugel  et al 1985; McCarthy et al 1987; Chambers, Miley \& van Breugel 1987, Dey et al 1997, Croft et al 2006) and theoretical arguments (e.g., De Young 1981,1989,  Rees 1989;  Begelman \& Cioffi 1989; Mellema, Kurk \& R\"ottgering 2002, Fragile et al 2004, Silk 2005) suggest that active galactic nuclei (AGN) 
can also induce the formation of stars in their host galaxies. The influence of the AGN may extend beyond mere enhancement of the star formation rate and create peculiar  stellar morphologies (e.g., Gaibler et al 2012), perhaps even in satellites of the  host galaxy (e.g., Efremov 2012). 

In the present paper we report the discovery, in a deep spectroscopic search for extended Ly$\alpha$ emission, of an extended asymmetric Ly$\alpha$ halo around the z=3.045 obscured
QSO J033238.76-275121.6.  Inspection of existing HST/ACS images shows spatial structure in the halo's Ly$\alpha$ profile coinciding in projection along
the slit with a peculiarly shaped, double-lobed
galaxy. Although the redshift of the galaxy is uncertain, we present arguments that suggest the  galaxy may be at the redshift of the QSO. We show that this would require a very young
stellar population, the age of which would be unlikely to be synchronized with the short-lived AGN activity if there were
no causal connection between the two, possibly in the form of QSO feedback triggering the formation of the stars in the external
galaxy. This scenario may also provide a plausible resolution to the problem posed by the divergent photometric redshifts suggested in the literature for the galaxy.

\section[]{Observations}

\subsection[]{Spectroscopy and Archival Imaging}

\begin{table*}
\scriptsize
 \centering
 \begin{minipage}{170mm}
  \caption{Properties of the  QSO (A) and the object (B\&C)}
  \begin{tabular}{@{}rllccccc}
\hline 
 ID  & z$_{\rm spec}$ & z$_{\rm phot}$& GOODS$^i$ & V$^i$  & B-V$^i$ \\
 \hline
A & 3.045$^{a}, 3.951^{b}$& 2.51$^c$, 1.46$^d$,  2.970$^e$, 1.663$^f$, 3.001$^g$  &J033238.76-275121.6     &26.32$\pm$0.05  &1.41$\pm$0.23&   \\
B\&C & 3.045$^{a}$ & 1.560$^h$, 2.310$^f$ & J033238.88-275119.5    & 26.55$\pm$0.09& 0.31$\pm$0.173&   \\
\hline
\end{tabular}
\\
comments: a) this paper; b) Vanzella et al 2010;  c) Zheng et al 2004; d) Treister et al 2006; e) Brusa et al 2009; f) Cardamone et al 2010; \\
g) Luo et al 2010;  h) Grazian et al 2006; i) Giavalisco et al 2004
\end{minipage}
\end{table*}

\begin{figure*}
\includegraphics[scale=.32,angle=0,keepaspectratio = true]{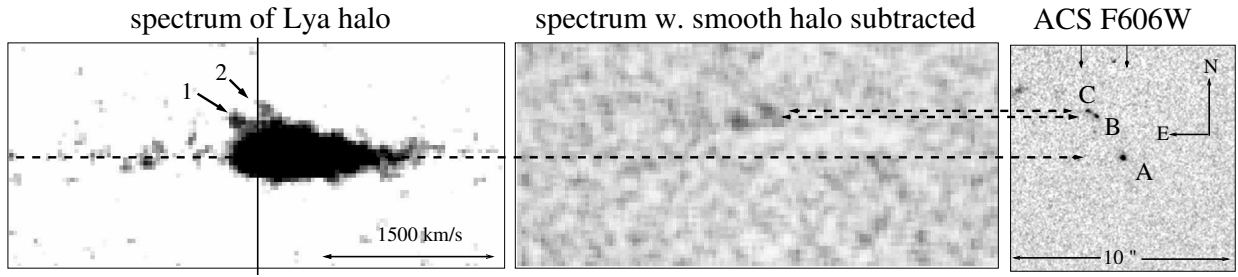}
\caption{LDSS3 spectrum of the Ly$\alpha$ emission line region (left); same spectrum, but with the smooth QSO halo subtracted (center),   and ACS F606W image  (right). The spatial size of the image and the extent
of the spectrum along the slit  shown here are 10".  A velocity scale of 1500km s$^{-1}$ is indicated in the spectrum. The dashed lines are meant to guide the eye and connect the position of the
QSO (bottom line, "A") and the continuum in the spectrum, and the two galaxy components "B" and "C" with the two knots 1 and 2 of Ly$\alpha$ emission, respectively. 
The vertical line in the spectrum approximately indicates the redshift 3.045 derived from the CIV emission line. The two downward arrows at the top of  the right panel indicate the approximate E-W positions of the slit jaws.\label{specplusim}}
\end{figure*}

The object in question is the third of several extended Ly$\alpha$ emitters in the long slit blind spectroscopic survey described in our earlier papers (Rauch et al 2011, paper I; Rauch et al 2013, paper II). A 2" wide long slit was placed for a total observing time
of 61.4 h on a random position centered on the Hubble Ultra Deep Field, with a position angle of 0 degrees. The slit happened to intersect at least partly the object known as  GOODS J033238.76-275121.6 (Giavalisco et al 2004)  or Combo-17 26280 (Wolf et  al 2001), which has
an absorption-corrected, rest-frame X-ray (0.5-8 keV) luminosity of $9.22\times10^{43}$
erg s$^{-1}$ (Xue et al 2011). Taken together with its hardness ratio and red optical colors this suggests classification as a type 2 AGN or obscured, borderline QSO
(e.g., Szokoly et al 2004; Padovani et al 2004).
A range of redshift estimates exist in the literature (see table 1),
including a spectroscopic redshift of z=3.951 (Vanzella et 2010). 
In our spectrum the QSO has a spatially asymmetric Ly$\alpha$ emission
line peaking near 4920 \AA, or at a redshift of 3.049. The presence and wavelength position of faint CIV emission suggests a  redshift 3.045, which is lower by about 300 kms$^{-1}$.  Lacking a good estimator of the systemic redshift we adopt the
CIV redshift z=3.045 (as the relation between the Ly$\alpha$ position and the systemic redshift is probably even more complicated). The precise redshift is immaterial for the main arguments in this paper.  At this redshift, the relative faintness in the observed B band (B-V=1.41), aside from a smaller correction by a factor 0.8 for absorption by the Lyman $\alpha$ forest, would be mainly due to reddening, consistent with a moderate amount of intrinsic extinction ($A_V\sim 0.7$) in the QSO's {\em rest frame} V band. With our new, lower redshift,
the flux measured for this object by Vanzella et al (2010) in a filter designed to capture escaping Lyman continuum radiation at $z\sim4$ would be explained by an overlap of the filter band with the QSO's (less strongly absorbed) non-ionizing continuum.

Even though the existing photometry shows the QSO to have a  V band magnitude of 26.31 (Giavalisco et al 2004), the continuum of the object is barely visible in our spectrum, suggesting that much of the light from the QSO may not have gone through  the 2" wide slit.

The spectral profile of the Ly$\alpha$ emission line peak is asymmetric, with a red shoulder, a steep drop going blueward, and some faint emission at the bluest end (fig.\ref{speccontour}).
The total flux transmitted through the slit is $(5.4\pm1.1)\times10^{-17}$ erg cm$^{-2}$s$^{-1}$. The emission line is spatially asymmetric as well, with the flux extending
further out to the N (top in fig. \ref{speccontour}). The lowest discernible flux density contour in the 2-d spectrum (fig.\ref{speccontour}), at $7\times10^{-20}$ erg cm$^{-2}$s$^{-1}$\AA$^{-1}$, extends out to about
17 proper kpc, encompassing two faint point-like emission peaks shown as "1" and "2" in fig. \ref{specplusim},
with individual Ly$\alpha$ fluxes  $(7.1\pm2.1)\times10^{-19}$ erg cm$^{-2}$s$^{-1}$ and $(7.8\pm1.5)\times10^{-19}$ erg cm$^{-2}$s$^{-1}$,
i.e., 3.5 and 5.2 $\sigma$ detections, respectively. These faint features can still be discerned separately in the two subsets of data obtained in 2008 and 2009, suggesting that they are not noise artifacts. Removing the smooth, diffuse halo with a  template determined
from the spatial profile redward of the two features reveals them as the only significant remaining features in the spectral
range shown (central panel in fig. \ref{specplusim}). 
The separation in wavelength space between 1 and 2 amounts to about 290 kms$^{-1}$, and they straddle the adopted "systemic" redshift for the QSO (vertical line in left panel of fig. \ref{specplusim}).
A comparison with the publicly available GOODS-S ACS F606W cutout images (Giavalisco et al 2004; 
fig. \ref{specplusim}, right panel) shows, in addition to the image of the QSO ("A"), two faint continuum emitters ("B" and "C") at approximately the same N-S angular 
distance $\theta$ from A ($\theta(B-A)=(1.83\pm0.03)$"; $\theta(C-A)=(2.11\pm0.03)$") that separates the objects 1 and 2 from the continuum trace of the QSO (($1.59\pm0.20$)" and
($2.16\pm0.20$)", respectively,  left panel). Objects B and C form an apparently  dumbbell-shaped structure in the ACS images. With similar broad band magnitudes in several optical filters, they may in fact be one, spatially coherent object. We shall refer to them as a single structure in what follows
but cannot rule out from the existing observations that they are two separate galaxies.

\subsection[]{The object  B\&C - unrelated foreground galaxy, or young starburst at the same redshift as the QSO ?}

The close alignment of  the B\&C continuum emission in projection with the faint Ly$\alpha$ emission 1 and 2 in the 2-d spectrum
may suggest a common origin in the same object. However, the two photometric redshifts for B\&C do not agree with the spectroscopic redshift
of the Ly$\alpha$ patches (and of the QSO; see table 1), nor do they agree among themselves.
Possible explanations for the discrepancy include the faintness of the object, the difficulty of spatially resolving the 
contributions from QSO and galaxy in ground-based observations, variable amounts of Ly$\alpha$ emission in the broad bands, 
and possibly the absence of templates from the photometric redshift determinations 
representing what may be an unusual stellar population.  In particular, the relatively blue colours of the object would be naturally
explained by the properties of a very young / peculiar stellar population at redshift 3.045, as described below. 
The photometry publicly available for constraining the rest-frame UV colours of the B\&C object(s) includes
HST ACS bands from original GOODS photometry by Giavalisco et al (2004), ground-based  VLT-VIMOS photometry of Grazian et al (2006),
and two additional ground-based U band observations from the ESO WFI (the so-called
U35 and U38 bands), which
were combined into single publicly available images by the GaBoDS group (Hildebrandt et al 2006). Cardamone et al (2010) have
measured U band fluxes for those images and included them in their catalogue.  Although the ground-based WFI images are noisy,
these bands are situated close
in wavelength to the Lyman limit of the Ly$\alpha$ halo and can at least in principle  constrain the redshift
of the B\&C system. 
For our analysis below we adopt these values in addition to the VIMOS U band and the ACS images.

\begin{figure*}
\includegraphics[scale=.45,angle=0,keepaspectratio = true]{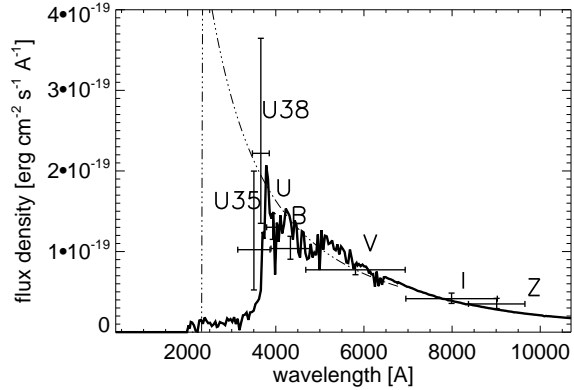}
\caption{Best fit to the broad band fluxes for galaxy B\&C with filters WFI U35, U38, U$_{\mathrm vimos}$ and HST ACS B,V,I, and Z,
The thick solid line shows a Starburst99 model spectrum (Leitherer et al. 1999) with total stellar mass $1.4\times10^7$ $M_\odot$
and an age of $2\times10^6$ yr, redshifted to z=3.0449 and attenuated by the average Ly$\alpha$ forest opacity blueward
of 1215.67 \AA\ in the rest frame.  The dash-dotted line is meant to illustrate a hypothetical flat spectrum ($\beta=-2$) source at z=2.310 (ignoring intergalactic absorption above the Lyman limit).\label{fluxmodel}}
\end{figure*}

For the galaxy to be at the redshift of the QSO, the steepness of the spectrum, even in the non-ionizing continuum, would require the spectrum to be dominated by young stars. The QSO could boost the continuum emitted by the galaxy
below the Lyman limit when part of the QSO's ionizing radiation is "reflected" by neutral hydrogen surrounding the galaxy, but the flux of the QSO seems
to fall short by considerable amounts for this to work. Moreover, unless the gas is optically thin, the effect would also produce a reflected,
proportional 2-photon continuum redward of Ly$\alpha$ that would tend to flatten the continuum slope.
We thus attempted to fit the broad-band optical measurements of objects B\&C with a young starburst spectrum at the redshift of the
QSO, using the library of starburst99 models (Leitherer et al 1999). Absorption by the Ly$\alpha$ forest was taken into account by using the "MC-Kim" transmission  of Bershady et al (1999). 

An instantaneous starburst
with the Starburst99 default standard initial mass function (IMF), an age of $2.0\times 10^6$ years, and a total stellar mass of $1.4\times10^7$M$_{\odot}$ fits the observed SED with a $\chi^2$ corresponding to an 11\% probability (with the total flux being the only free parameter). More extreme stellar populations
can in principle improve the fit further. Choosing an extreme IMF that is flat or consists of only massive stars $(M>5 M_{\odot})$ and an age of the burst of only $1.4\times10^6$ years would increase 
the flux near the Lyman limit further and produce a $\chi^2$ probability of about 30\%, but may not be warranted on physical grounds.

Fig.\ref{fluxmodel} shows a comparison between the observed fluxes and the best-fit model spectrum. For illustrative purposes, the dashed-dotted line also shows
a flat (in $f_{\nu}$) spectrum  at lower redshift (z=2.310) (ignoring Ly$\alpha$ forest absorption), as if the object were at the photometric redshift derived by Cardamone et al (2010).
From this comparison it is not surprising that a large uncertainty in the U-band flux, with filters straddling the Lyman limit, can lead to a degeneracy in redshift. 
We conclude that the observations of the B\&C object are consistent with it being a galaxy at the redshift of the QSO if we assume the presence of a very young population of stars.

We do not know the lifetime of the QSO, but can estimate crudely the possibility of finding two "young objects" (the QSO,
and object B\&C) as close together as observed. 
The probability distribution of a chance coincidence of a second object with a magnitude $V<27$ and a $B-V<0.35$ (as observed for B\&C) to occur near a similarly "blue" 
pre-selected object, estimated from random draws from the Giavalisco et al (2004) galaxy catalogue, makes the present arrangement  approximately a 3.4 $\sigma$
excursion. More severe upper limits on the chance probability of finding a (short-lived) AGN and a short-lived population of young stars
can be derived from the respective duty cycles.  Assuming a duty cycle of a few percent for QSOs (Yu \& Tremaine 2002; Nandra et al 2002) and $\sim 6\%$ for starbursts in dwarf galaxies (Lee et al 2009), gives a sub-percent probability for
finding both in the active stage. Thus, if the galaxy indeed has a young stellar
population, its formation must be closely coeval with the QSO activity.

\section[]{The origin of the Ly$\alpha$ emission}

Producing the observed Ly$\alpha$ fluxes in the features 1 and 2 outside the main QSO emission line through recombination following photoionization 
requires an ionization rate
\begin{eqnarray}
\dot{N}^{\rm ion} = \frac{3}{2}\left(\frac{F_{{\rm Ly}\alpha}}{{\rm h}\nu_{{\rm Ly}\alpha}}\right)4\pi D_L^2 = 1.2\times10^{52}\left(\frac{F_{{\rm Ly}\alpha}}{1.5\times10^{-18}}\right)\ {\rm s}^{-1}, 
\end{eqnarray}
where a case B conversion from ionizing photons into Ly$\alpha$ photons has been assumed, and the combined Ly$\alpha$ flux $F_{{\rm Ly}\alpha}$ of objects 1 and 2 is $(1.5\pm 0.3)\times 10^{-18}$ erg cm$^{-2}$s$^{-1}$. $D_L$ is the luminosity distance at z=3.045.
The stellar population of object B\&C, at a magnitude $m_{\rm AB}$ = 26.6 in the observed V band, produces HI ionizing photons at a rate of approximately $1.55\times 10^{53}$ s$^{-1}$ (using the
assumptions made in Paper I). Thus, the supply of ionizing photons, if completely trapped in the gaseous halos of B\&C  and converted into Ly$\alpha$ as assumed above,
would exceed the number of ionizations required to explain the observed Ly$\alpha$  emission by a comfortable factor 13 (ignoring
slit losses for Ly$\alpha$).

Alternatively,  the Ly$\alpha$  emission from objects 1 and 2  could,
in principle, also be powered by the QSO. 
The spatially extended, diffuse
Ly$\alpha$  emission centered on the QSO extends more than half the 
distance between the QSO and objects 1 and 2   at a similar surface
brightness level. This suggests that the ionizing flux from  the QSO
is at least in some directions of the same order of magnitude as required to power  
the Ly$\alpha$  emission of objects 1 and 2.

\section[]{Conclusions}

The asymmetric extent of the Ly$\alpha$ halo is common for "fuzz" around QSOs (e.g., Christensen et al 2006),
presumably reflecting the tilt of the ionization cone away from the line of sight to the observer. We have argued that the 
extension of the halo in the direction of the B\&C object and the alignment of the spatial substructure in the Ly$\alpha$ emission
with the broad-band imaging positions of the galaxy
suggest a physical proximity and possible causal connection between the QSO and the object. Coeval activity is further suggested by the coincidence between apparently  recent star formation  and the active state of the QSO.
The similarity of the lobes of B \& C in brightness and colours  suggests that they may constitute a single coherent galaxy, with the velocity difference of 290 kms$^{-1}$ perhaps being due to rotation. The appearance of two spatially distinct
Ly$\alpha$ emission regions for this object as opposed to a common halo may simply reflect the two discrete stellar regions.
Alternatively, the separate Ly$\alpha$ peaks may 
be caused when the Ly$\alpha$ emission gets suppressed in the center of the B\&C structure either by the QSO physically expelling the gas, or by largely ionizing it. In either case, any further ionizing radiation may
just pass through the cleared patch and not be available for conversion to Ly$\alpha$. As for 
the hole in the stellar population, it is not clear how this could be produced after the stars are already in place, but it does appear more likely that the QSO may determine in the first place where the stars are going to form (and where not). If feedback from the QSO is influencing the properties of this object, then the fact that the structure is largely symmetric must either be a fortunate coincidence
(e.g., a chance alignment with whatever symmetry the QSO feedback mechanism possesses; such as a jet that just happens to hit the galaxy at the right spot), or the structure with its stars and gas distribution must be formed as a direct
result of the QSO feedback. In the latter case, the star-formation geometry naturally reflects the directional symmetry of the feedback. Such a possibility has been suggested  in the literature (see introduction), both to explain
the observational alignments of star-forming regions and outflows from radiogalaxies, and as a theoretically predicted consequence of AGN feedback. It appears that the current
case may be a high-redshift candidate case for AGN-induced star-formation in an external galaxy.  This interpretation, if confirmed,  would be consistent with the idea discussed  in papers I and II, that asymmetric, extended Ly$\alpha$ emitters
trace a crucial phase in the galaxy formation process, where interactions among the building blocks of future 
normal galaxies influence the gas dynamics, star formation process, and production of ionizing radiation in these objects.

\section*{Acknowledgments}

We acknowledge helpful discussions with Carolin Cardamone, Andrea Grazian, and Francois Schweizer.  We  thank the staff of the Las Campanas Observatory  for their help with the observations.
MR is grateful to the IoA in Cambridge and to the Raymond and
Beverley Sackler Distinguished Visitor program for hospitality and support in summer 2011 and 2012, when some of this work was done, and
to the National Science
Foundation for grant AST-1108815.  GB has been supported by the Kavli Foundation. JRG acknowledges  a Millikan Fellowship at Caltech. We acknowledge the use of the extinction calculator by Doug Welch
\footnote{http://dogwood.physics.mcmaster.ca/Acurve.html}.







\bsp

\label{lastpage}


\begin{thebibliography}{}

	
	

	
\bibitem[Begelman \& Cioffi 1989]{beg89} Begelman, M. C., Cioffi, D. F., 1989, ApJ, 345, 21


\bibitem[Bershady, Charlton \& Geoffroy 1999]{ber99}Bershady, M. A. Charlton, Jane C., Geoffroy, J. M., 1999, ApJ, 518, 103

\bibitem[Brusa et al 2009]{bru09}Brusa, M., Comastri, A., Gilli, R., Hasinger, G., Iwasawa, K., Mainieri, V., Mignoli, M., Salvato, M., Zamorani, G., Bongiorno, A., and 12 coauthors,  2009, ApJ, 693,8	

\bibitem[Cardamone et al 2010]{car10}Cardamone, C. N., van Dokkum, P. G., Urry, C. M., Taniguchi, Y., Gawiser, E., Brammer, G., Taylor, E., Damen, M., Treister, E., Cobb, B. E., and 6 coauthors, 2010, ApJS, 189, 270	

\bibitem[Chambers et al 1987]{cha87}Chambers, K. C., Miley, G. K., van Breugel, W., 1987, Nature, 329, 604	
    
\bibitem[Christensen et al 2006]{chr06} Christensen, L., Jahnke, K., Wisotzki, L., Sanchez, S. F., 2006, A\&A, 459, 717

\bibitem[Croft et al 2006]{cro06} Croft, S.,  van Breugel, W., de Vries, W.  Dopita, M., Martin, C., Morganti, R., Neff, S., Oosterloo, T., Schiminovich, D.,  Stanford, S. A., van Gorkom, J.,  2006, ApJ, 647, 1040

\bibitem[Dey et al 1997]{dey97}Dey, A., van Breugel, W., Vacca, W. D., Antonucci, R., 1997, ApJ, 490, 698	

\bibitem[De Young 1981]{dey81}De Young, D. S., 1981, Nature 293, 43 

\bibitem[De Young 1989]{dey89}De Young, D. S., 1989, ApJ, 342, L59

\bibitem[Efremov 2012]{efr12}Efremov, Y.N., 2013, MNRAS, 429L, 75

\bibitem[Fabian 1999]{fab99}Fabian, A.C., 1999, MNRAS, 308, 39

\bibitem[Farrah et al 2012]{far12} Farrah, D., Urrutia, T., Lacy, M. Efstathiou, A., Afonso, J., Coppin, K., Hall, P. B. Lonsdale, C., Jarrett, T., Bridge, C., Borys, C., Petty, S., 2012, ApJ, 745, 178

\bibitem[Fragile et al 2005]{fra05}Fragile, P. C., Murray, S. D., Anninos, P., van Breugel, W., 2004, ApJ, 604, 74

\bibitem[Gaibler et al 2012]{gai12} Gaibler, V., Khochfar, S., Krause, M., Silk, J., 2012, MNRAS, 425, 438

\bibitem[Giavalisco et a 2004]{giv04} Giavalisco, M. and the GOODS Team,  2004, ApJ, 600, L93

\bibitem[Grazian et al  2006]{gra06} Grazian, A., Fontana, A., de Santis, C., Nonino, M., Salimbeni, S., Giallongo, E., Cristiani, S., Gallozzi, S., Vanzella, E., 2006, A\&A, 449, 951

\bibitem[Haiman \& Rees 2001]{hai01} Haiman, Z., Rees, M.J., 2001, ApJ, 556, 87	

\bibitem[Hildebrandt et al 2006]{hil06}Hildebrandt, H., Erben, T., Dietrich, J. P., Cordes, O., Haberzettl, L., Hetterscheidt, M., Schirmer, M., Schmithuesen, O., Schneider, P., Simon, P., Trachternach, C., 2006, A \& A, 452, 1121


\bibitem[Lee et a 2009]{lee09}Lee, J. C., Kennicutt, R. C. jr., Funes, S. J.,
Jose, G., Sakai, S., Akiyama, S., 2009, ApJ, 692, 1305

\bibitem[Leitherer et al 1999]{lei99} Leitherer, C., Schaerer, D., Goldader, J. D., Gonzalez Delgado, R. M., Robert, C., Kune, D. F., de Mello, D. F., Devost, D., Heckman, T. M., 1999, ApJS, 123, 3

\bibitem[Luo et al 2010]{luo10} Luo, B., Brandt, W. N., Xue, Y. Q., Brusa, M., Alexander, D. M., Bauer, F. E., Comastri, A., Koekemoer, A., Lehmer, B. D., Mainieri, V., and 4 coauthors, 2010, ApJS, 187, 560

\bibitem[Maiolino et al 2012]{mai12} Maiolino, R., Gallerani, S., Neri, R., Cicone, C., Ferrara, A., Genzel, R., Lutz, D., Sturm, E., Tacconi, L. J., Walter, F., Feruglio, C., Fiore, F., Piconcelli, E., 2012 MNRAS, 425, 66

\bibitem[McCarthy et al 1987]{mcc87}McCarthy, P. J., van Breugel, W., Spinrad, H.,  Djorgovski, S., 1987, ApJ, 321, 29

\bibitem[Mellema, Kurk \& R\"ottgering 2002]{mel02}Mellema, G., Kurk, R\"ottgering, H. J. A., 2002, A\&A, 395, L13

\bibitem[Nandra et al 2002]{nan02} Nandra, K., Mushotzky, R. F., Arnaud, K., Steidel, C. C., Adelberger, K. L., Gardner, J. P.,
Teplitz, H. I., Windhorst, R. A., 2002, ApJ, 576, 625

\bibitem[Padovani et al 2004]{pad04}Padovani, P., Allen, M.G., Rosati, P., Walton, N.A., 2004, A\&A, 424, 545

\bibitem[Rauch et al 2011]{rau11b}Rauch, M. Becker, G. D., Haehnelt, M. G., Gauthier, J.-R., Ravindranath, S., Sargent, W. L. W.,2011, MNRAS, 418, 1115 (Paper I)

\bibitem[Rauch et al 2012]{rau12}Rauch, M. Becker, G. D., Haehnelt, M. G., Gauthier, J.-R., Sargent, W. L. W., 2013, MNRAS, 429, 429 (Paper II)

\bibitem[Rees 1989]{ree89} Rees, M.J., 1989, MNRAS, 231, 1

\bibitem[Schawinsky et al 2009]{sch09} Schawinski, K., Virani, S., Simmons, B., Urry, C. M., Treister, E., Kaviraj, S., Kushkuley, B., 2009, ApJ, 692, 19

\bibitem[Silk \& Rees 1998]{sil98} Silk, J., Rees, M. J., 1998, A\&A, 331, 1

\bibitem[Silk 2005]{sil05}Silk, J., 2005, MNRAS, 364, 1337


\bibitem[Szokoly et al 2004]{szo04}Szokoly, G. P., Bergeron, J., Hasinger, G., Lehmann, I., Kewley, L., Mainieri, V., Nonino, M., Rosati, P., Giacconi, R., Gilli, R., and 8 coauthors,  2004, ApJS, 155, 271
   
\bibitem[Treister et al 2006]{tre06}Treister, E., Urry, C. M. Van Duyne, J., Dickinson, M., Chary, R.-R., Alexander, D. M., Bauer, F., Natarajan, P., Lira, P., Grogin, N. A., 2006, ApJ, 640, 603

\bibitem[van Breugel et al 1985]{van85}van Breugel, W., Filippenko, A. V., Heckman, T., Miley, G., 1985,ApJ, 293, 83	


\bibitem[Vanzella et al 2010]{van10}Vanzella, E., Giavalisco, M., Inoue, A. K., Nonino, M., Fontanot, F., Cristiani, S., Grazian, A., Dickinson, M.,  Stern, D., Tozzi, P., and 7 coauthors, 2010, ApJ, 725 1011	


\bibitem[Wolf et al 2001]{wol01} Wolf, C.,  Dye, S., Kleinheinrich, M.,  Meisenheimer, K., Rix,. H.-W.,  Wisotzki, L., 2001, A\&A, 377, 442
  
\bibitem[Xue et al 2011]{xue11}Xue, Y. Q., Luo, B., Brandt, W. N., Bauer, F. E., Lehmer, B. D., Broos, P. S., Schneider, D. P., Alexander, D. M., Brusa, M., Comastri, A., and 15 coauthors, 2011, ApJS, 195, 10

\bibitem[Yu \& Tremaine 2002]{yut02}Yu, Q.,  Tremaine, S., 2002, MNRAS, 335, 965

\bibitem[Zheng et al 2004]{zhe04}Zheng, W., Mikles, V. J., Mainieri, V., Hasinger, G., Rosati, P., Wolf, C., Norman, C., Szokoly, G., Gilli, R., Tozzi, P., and 3 coauthors, 2004, ApJS, 155, 73


\end{thebibliography}
\end{document}